\documentstyle[aps]{revtex}

\begin{document}
\twocolumn[\hsize\textwidth\columnwidth\hsize
\csname@twocolumnfalse%
\endcsname


 \draft

  \title{
  Surface Region of Superfluid Helium as an Inhomogeneous
Bose-Condensed
  Gas}

  \author{A. Griffin\cite{ag} and S. Stringari}
  \address{Dipartimento di Fisica and INFM, Universit\`{a} di
Trento,
  I-38050 Povo, Italy}

  \date{submitted to PRL, May 17, 1995}

  \maketitle

  \begin{abstract}
  We present arguments that the low density surface region of
  self-bounded superfluid $^4$He systems is an inhomogeneous
  dilute Bose gas, with
  almost all of the atoms occupying the same single-particle state
at
  $T = 0$.
  Numerical evidence for this complete Bose-Einstein condensation
  was first given by the many-body variational calculations of
$^4$He
  droplets by Lewart, Pandharipande and Pieper in 1988.  We show
that
the low
  density
  surface region
  can be treated rigorously using a generalized Gross-Pitaevskii
  equation for the Bose order parameter.

  \end{abstract}

]

 \narrowtext

  In recent years, there has been a major effort to produce
  atomic-like gases in a Bose-condensed phase, using excitons,
  spin-polarized Hydrogen and laser-trapped alkali atoms (for
reviews
  of
  this work, see Refs.~\protect\cite{{ag-ds-ss95},{gt94}}).  In
such
  systems, the fraction of atoms in the lowest energy
single-particle
  state could approach 100\% at $T=0$. This
  unique phase of matter was first studied by Einstein and London
and
  would exhibit many unusual
  properties \protect\cite{gt94}, including superfluidity.
  In contrast, in bulk superfluid $^4$He, the
  condensate fraction is
  only about 10\% at $T=0$ \protect\cite{ag-ds-ss95,ag93}.
  Thus the ground state wavefunction does not have the simplicity
of
  a gas with complete Bose-Einstein condensation (BEC).

  In the present letter, we argue that such a dilute Bose-condensed
gas
   is in fact already present in the
  surface region of superfluid $^4$He, where the density becomes
very
  small.  This striking phenomenon has apparently been overlooked
in
  the extensive theoretical literature on surfaces, films and
  droplets of superfluid $^4$He (for references, see
  Ref.~\protect\cite{fd-al-etc}). Moreover, in
  this low density surface region, where one has almost 100\% BEC
  into a single state, one can do
  microscopic calculations of the order parameter using
  the  standard field-theoretic description of an inhomogeneous
  Bose-condensed fluid \protect\cite
  {alf-jdw71,pch-pcm65}. Our discussion is limited to
  ground state
  properties  $(T = 0)$ but we make a few remarks about finite
  temperature at the end.

  Striking confirmation of the preceding physical argument is
  available in the results of \mbox{Lewart}, Pandharipande and
Pieper
  (LPP)
  \protect\cite{dsl-vkp-scp88}, who studied $^4$He droplets (with
up
  to 240 atoms) using a variational ground-state wavefunction
  approach \protect\cite{vrp-scp-rb86}.  These authors calculated
the
  single-particle ``natural
  orbitals'' $\psi_{i}({\bf r})$ which diagonalize the
  single-particle
  density matrix, giving
  \begin{equation}
  \rho_{1}({\bf r},{\bf r}^{\prime})=\sum_i n_i\psi_{i}^{\ast}
  ({\bf r})\psi_{i}({\bf r}^{\prime}),
  \label{eq1}
  \end{equation}
  where $n_i$ is the occupation probability of the state
  $\psi_{i}({\bf r})$.
  The local density is given by $\rho({\bf r})=\rho_{1}({\bf
r},{\bf
  r})$.  By direct
  numerical calculation, LPP found that the
 node-less $1s$-state $\psi_{1s}({\bf r})$ had an occupation
  probability of
  $n_{1s}\simeq 25$, while for all the other states, $n_i < 0.5$.
  They identified (and justified) this $1s$ state as the
  Bose-condensed state, with an associated local condensate density
  being
  given by
  \begin{equation}
  \rho_{c}({\bf r})=n_{1s}\vert\psi_{1s}({\bf r})\vert^2\ .
  \label{eq2}
  \end{equation}
  LPP evaluated the
  total local density $\rho({\bf r})$ and the condensate local
  density
  $\rho_{c}({\bf r})$ using (\protect\ref{eq2}).  Their results for
  $N=70$ atoms are replotted
  in
  Fig.~\protect\ref{fig1}.  However, LPP
  did not explicitly discuss the {\it implications} of the most
  striking
  feature shown in Fig.~\protect\ref{fig1}, namely that since in
the
  surface
  region almost all the atoms at a given value of $r$ are
  in the $\psi_{1s}({\bf r})$ state, this
  surface region
  provides a physical realization of the long sought-for
  dilute Bose-condensed gas \protect\cite{ag-ds-ss95,gt94}.

  An important aspect of the calculations in
  Ref.~\protect\cite{dsl-vkp-scp88} is that near the
  center of the drop, $\rho({\bf r})$ and
  $\rho_{c}({\bf r})$ are very close to the values in bulk
  superfluid
  $^4$He, with $\rho_{c}({\bf r}) \simeq 0.1\rho({\bf r})$.  This
  indicates that even
  droplets of $N=70$ atoms are large enough to ensure that the
  surface region density profile will be very similar to that of a
  planar
  free surface of bulk superfluid $^4$He.  This is confirmed by
more
  recent calculations  on larger droplets \protect\cite{sac93}. In
  the
  case of planar (vs spherical) symmetry, the atoms will Bose
  condense
  into states with zero momentum parallel to the surface
  \protect\cite{ek85}.

  Krotscheck \protect\cite{ek85} first found that the local
  condensate
  density $\rho_{c}({\bf r})$ increased when the total density
  $\rho({\bf r})$
  decreased in
  liquid $^4$He in contact with a wall.  Campbell
  \protect\cite{cec93} has recently drawn attention to the results
in
  Ref.~\protect\cite{dsl-vkp-scp88,vrp-scp-rb86,sac93,ek85} and
  discussed them in the context of the formal theory of
  off-diagonal
  long range order (ODLRO) in liquid $^4$He droplets.
  Apart from these papers,
  Bose-condensation has been ignored in the literature on
  inhomogeneous Bose liquids at $T=0$ (for
  references, see \protect\cite{fd-al-etc} and
  \protect\cite{ek-gxq-wk85}). Moreover, even in
Refs.~\protect\cite
  {dsl-vkp-scp88,vrp-scp-rb86,sac93,ek85,cec93}, what is still
  missing
  is a physical interpretation of the numerical results for
  $\rho_{c}({\bf r})$ at surfaces
  as evidence for (and relevance of) a low density Bose gas in
which
  the condensate fraction can be 100\%, and the important
  {\it implications} of this new picture.  It is this aspect which
  the
  present letter addresses.

  The numerical results in Fig.~\protect\ref{fig1} are consistent
  with the empirical formula given in Ref.~\protect\cite
  {dsl-vkp-scp88},

  \begin{equation}
  \rho_{c}(r) =\rho(r)\left[1-0.68
  \frac{\rho(r)}{\rho_B}\right ]^2\ ,
  \label{eq3}
  \end{equation}
  where $\rho_B$ is the bulk or saturation density of superfluid
  $^4$He at $T=0$. It is easy to verify using
  (\protect\ref{eq3}) that the local condensate {\it fraction}
  $\rho_c( r)/\rho( r)$ smoothly increases from 0.1 to
  unity as $\rho( r)$ goes from $\rho_B$ to zero.
    One
  of the most interesting features shown by Fig.~\protect\ref{fig1}
  is that $\rho_c(r)$ develops a maximum.  Using
(\protect\ref{eq3}),
  this occurs when $\rho(r) \simeq 0.5 \rho_B$ and
  corresponds
  to $\rho_c(r) \simeq 0.22\rho_B$.  Thus the results of
  Ref.~\protect\cite{dsl-vkp-scp88} predict that in a certain
region,
  the density of
  $^4$He atoms which are Bose-condensed can be {\it twice} as large
  as the
  bulk value $\rho_c\simeq 0.1\rho_B$!   Of course,
  as $\rho(r)$
  decreases to zero, so must $\rho_c(r)$.   While the precise
magnitude may depend on the approximate calculations used
in \protect\cite{dsl-vkp-scp88}, we believe that the {\it
  increase} in $\rho_c(r)$ as we go from the center of the $^4$He
  drop is a real effect. As noted in
  Ref.~\protect\cite{dsl-vkp-scp88}, the initial increase in
  $\rho_c( r)$
  shown in Fig.~\protect\ref{fig1} for $\rho( r) < \rho_B$
  is consistent with the calculated and observed decrease in
$\rho_c$
  in bulk liquid $^4$He under pressure \protect\cite{ss91}.  Thus
in
  superfluid $^4$He,
  $\partial\rho_c(\rho)/\partial\rho \simeq -0.3$ for $\rho$
  close to $\rho_B$.  In contrast, in a dilute hard sphere Bose
gas,
  $\partial\rho_c(\rho)/\partial\rho \lesssim 1$
  \protect\cite{alf-jdw71}.

  Independently of the numerical results in
  Refs.~\protect\cite{dsl-vkp-scp88,sac93}, it is easy to see that
  the
  general structure of the variational many-particle wavefunction
  which LPP and others \protect\cite
  {vrp-scp-rb86,sac93,rmb70,ccc-mc73,ksl-mhk-gvc75} have used to
  treat
  the free
  surface of liquid $^4$He already builds in the essential physics
of
  a
  surface region which is completely Bose-condensed into one
  single-particle state.  In their simplest form, such
wavefunctions
  are assumed to have the Feenberg form
  \begin{equation}
  \Psi({\bf r}_1,\ldots{\bf r}_N) = Ae^{[-
  \frac{1}{2}\sum\limits_{i<j}u({\bf r}_i-{\bf r}_j)
  -\frac{1}{2}\sum\limits_i
  t({\bf r}_i) ]}\ ,
  \label{eq4}
  \end{equation}
  including one and two-particle correlations. The function
  $t({\bf r})$
  controls the shape of the density profile $\rho({\bf r})$
  associated with this wavefunction.  It is
  easy
  to see that (\protect\ref{eq4}) can be rewritten in the form
  \begin{equation}
  \Psi({\bf r}_1,\ldots{\bf r}_N)=A\Psi_{{\rm
  Jastrow}}({\bf r}_1,\ldots{\bf r}_N)\Psi_{{\rm
  gas}}({\bf r}_1,\ldots r_N)\ ,
  \label{eq5}
  \end{equation}
  where $\Psi_{{\rm Jastrow}}$ alone describes a bulk liquid and
  \begin{equation}
  \Psi_{{\rm gas}}({\bf r}_1,\ldots{\bf r}_N)
  =\prod^N_{i=1}\phi_0({\bf r}_i)
  \label{eq6}
  \end{equation}
  describes a Bose gas of $N$ atoms all occupying the same
  single-particle state given by $\phi_0({\bf
  r})=e^{-\frac{1}{2}t({\bf r})}$
  ({\it
  i.e.}, with 100\% BEC). Choosing $u({\bf r})$ and $t({\bf r})$
  appropriately, wavefunctions such as
  (\protect\ref{eq4}) allow one the variational freedom to describe
  atoms in
  the bulk region (where
  $\Psi_{{\rm gas}} \simeq 1$) and in the low density surface
region
  (note that when the atoms are far apart, one has $\Psi_{{\rm
  Jastrow}}\simeq 1$).  It is important to
  realize that
  $\phi_0({\bf r})$ is not the single-particle
  ``natural
  orbital'' in (\protect\ref{eq2}), since the latter is defined
relative to the complete wavefunction in (\protect\ref{eq4}).
  The interpretation of (\protect\ref{eq4}) as
  described above in (\protect\ref{eq5}) and
  (\protect\ref{eq6}) has not been emphasized in the
  previous literature on superfluid surfaces (see, however, Ref.
  \protect\cite{dsl-vkp-scp88}) but we think the present analysis
  clarifies the physics behind the complete BEC found in the
surface region.

   In the rest of this letter, we discuss how our new
  picture can be used to describe the surface region in a rigorous
  fashion using the field-theoretic analysis of an inhomogeneous
  Bose-condensed fluid \protect\cite{ag93,alf-jdw71}.  In addition,
  we point out that this has
  important implications concerning density functional
  approaches used to calculate the surface properties of superfluid
  $^4$He \protect\cite{fd-al-etc,ss-jt87,jdr-mh-etc}.

  The field-theoretical approach to Bose
  systems is based on the presence of the symmetry-breaking order
  parameter $\Phi ({\bf r}) \equiv \langle \hat \psi({\bf r})
  \rangle$
  as the anomalous average of the field
  operator $\hat\psi({\bf r})$ (see Ch.~3 of Ref.
  \protect\cite{ag93}). The
  local
  condensate density is given by $\rho_{c}({\bf r})=\vert\Phi({\bf
  r})\vert^2$.
  The equation for the order parameter can be obtained starting
from
  the Heisenberg equation for the field operator
  \begin{equation}
  -i\frac{\partial}{\partial t} \hat\psi({\bf r})
  = [H^{\prime},\hat\psi({\bf r})]
  \label{eq7}
  \end{equation}
  where $H^{\prime} = H - \mu N$ is the grand canonical
Hamiltonian.
  By carrying out explicitly the commutator of (\protect\ref{eq7})
  and taking the statistical average on the equilibrium state of
the
  system one obtains the exact equation
\protect\cite{pch-pcm65,pn-dp90}
  \begin{equation}
  \left [ -\frac{\hbar^2\nabla^2}{2m}-\mu\right ]\Phi({\bf r})
+\int
  d{\bf r}^\prime v({\bf r} - {\bf r}^\prime)\langle
\hat\psi^\dagger
  ({\bf r}^\prime)\hat\psi({\bf r}^\prime)\hat\psi({\bf
r})\rangle=0\ ,
  \label{eq8}
  \end{equation}
  where $v({\bf r}-{\bf r}^{\prime})$ is the bare He-He interatomic
  potential. When applied to the homogeneous liquid,
(\protect\ref{eq8})
  provides an exact, non-trivial relationship between the chemical
  potential $\mu$ and the long range behaviour of the non-diagonal
  two-body density matrix of a Bose condensed system
\protect\cite{notemu}.

  In the case of the free surface, the value of the chemical
potential coincides with the bulk $^4$He binding energy
  ($-7.15$K at $T=0$). In this case, (\protect\ref{eq8}) can be
used to
  investigate the behaviour of the order parameter $\Phi({\bf r})$
when ${\bf r}$ is in the
asymptotic, low
  density region far from the surface.
Our reasoning is as follows. The dominant contribution to the
integral in (\protect\ref{eq8}) comes from ${\bf r}^\prime$ in the
bulk region, where $\rho({\bf r}^\prime)$ is large. For this
contribution, $\vert{\bf r}-{\bf r}^\prime\vert$ is large and the
correlation function in (\protect\ref{eq8}) is given by the
asymptotically exact formula
  \begin{equation}
  \langle \hat\psi^\dagger ({\bf r}^\prime)\hat \psi({\bf
r}^\prime)
  \hat\psi({\bf r})\rangle\vert_{\vert{\bf r}-{\bf r}^{\prime}\vert
\to
  \infty} = \langle \hat\psi^\dagger ({\bf r}^\prime)\hat \psi({\bf
r}^\prime)
  \rangle \langle \hat\psi({\bf r})\rangle = \rho({\bf
r}^\prime)\Phi({\bf r})
  \ .
  \label{eq9}
  \end{equation}
  This may be viewed as a generalization of the
  Penrose--Onsager result  \protect\cite{op-lo56}
  \begin{equation}
  \langle \hat\psi^\dagger ({\bf r}^\prime) \hat\psi({\bf
r})\rangle
  \vert_{\vert{\bf r}-{\bf r}^{\prime}\vert \to
  \infty} = \langle \hat\psi^\dagger ({\bf r}^\prime) \rangle
\langle
  \hat\psi({\bf r})\rangle = \Phi^*({\bf r}^\prime)\Phi({\bf r})\
{}.
  \label{eq10}
  \end{equation}
Using (\protect\ref{eq9}) in (\protect\ref{eq8}), we conclude that
for ${\bf r}$ in the low density region, (\protect\ref{eq8})
reduces to
\begin{equation}
  \left\{-\frac{\hbar^2\nabla^2}{2m}-\mu +\int d{\bf r}^\prime
v({\bf
  r} -{\bf r}^\prime) \rho({\bf r}^\prime)\right\} \Phi({\bf r})=0\
,
  \label{eq11}
  \end{equation}
where the Hartree potential
\begin{equation}
  v_H({\bf r}) =  \int d{\bf r}^\prime v({\bf r} -
  {\bf r}^\prime) \rho({\bf r}^\prime)
  \label{eq12}
  \end{equation}
describes the field at ${\bf r}$ due to the bulk region where the
density is large. For such contributions, where $\vert{\bf r}-{\bf
r}^\prime\vert$ is large, only the long-range attractive van der
Waals' tail of $v({\bf r}-{\bf r}^\prime)$ is important. For
contributions to (\protect\ref{eq8}) when $\vert{\bf r}-{\bf
r}^\prime\vert$ is small ({\it i.\ e.\ }, when ${\bf r}^\prime$ is
in the low
surface region), one can use $\langle\hat\psi^\dagger({\bf
r}^\prime)\hat\psi({\bf r}^\prime)\hat\psi({\bf
r})\rangle\simeq\vert\Phi({\bf r}^\prime)\vert ^2\Phi({\bf r})$
since we are dealing with a region of complete BEC. Such
contributions are precisely those kept in the standard Gross-
Pitaevskii theory \protect\cite{epg61,alf-jdw71} and as usual one
needs to include additional multiple scattering terms
\protect\cite{alf-jdw71} which screen the hard core in $v({\bf r}-
{\bf r}^\prime)$ in (\ref{eq8}). However the details are not
important here (see
Ref.~\protect\cite{newref23}) since the low density at ${\bf r}$
makes these contributions to (\protect\ref{eq8}) negligible
compared to the ones discussed above. Similarly, while the
many-body screening of the hard core is not explicitly included in
(\protect\ref{eq11}) and (\protect\ref{eq12}), contributions to
$v_H({\bf r})$ when $\vert{\bf r}- {\bf r}^\prime\vert$ is small
are negligible because of the low density at ${\bf r}$.

In summary, we conclude that (\protect\ref{eq11}) allows one to
determine the exact asymptotic behavior of $\Phi({\bf r})$ far from
the surface of superfluid $^4$He \protect\cite{newref23}.
Interpreting (\protect\ref{eq11}) in physical terms, we see that
the bulk of the liquid produces an effective
  field which acts to stabilize the dilute inhomogeneous gas
  in the surface region.  The fact that the atoms in the low
density
  region only feel this effective field but otherwise are not
affected by
  their interactions explains why this region is fully
Bose-condensed.
  Equations with the same Hartree-like
  structure as (\protect\ref{eq11}) arise in discussions
  of the binding energy of an impurity atom on the free surface of
liquid $^4$He,
  based on
  variational wavefunctions \protect\cite{ibm-doe79}.

  For a planar surface centered at $z = 0$, the asymptotic behavior
of
  (\protect\ref{eq12}) reduces at large $z$ to $v_{H}(z) = -\rho
  _{B}
  \alpha/z^{3}$, where $\alpha$ only depends on the parameters of
the  He-He
  attractive potential \protect\cite{ibm-doe79}. Using this in
  (\protect\ref{eq11}),
  one finds that the order parameter has the
  form
  \begin{equation}
  \Phi (z)\ \propto  e^{-(Az+B/z^2)},
  \label{eq13}
  \end{equation}
  where $A = \sqrt{\frac{2m\vert\mu\vert}{\hbar^2}} \simeq
1$\AA$^{-1}$
  and $B = m \rho _{B} \alpha/2 \hbar^{2} A \simeq 5$\AA$^{2}$.
  The above discussion then suggests that  the
  exponential decay \protect\cite{notetr}
  exhibited by the resulting density profile $\rho
  (z) = \rho_{c}(z) = \vert \Phi(z) \vert^{2}$ is a direct
  consequence
  of the complete BEC in the low density region, which occurs (see
  Fig.~1) for $\rho (z) \lesssim 0.1 \rho_{B}$. This behavior
  differs, for example, from the surface density profile of a
  Fermi liquid, where one
  finds a faster decay, of the form $\rho(z)\ \propto\ z^{-2}e^{-2Az}$
  \protect\cite{notess}.
  Integrating
  (\protect\ref{eq13}) to find the total number of atoms in the
  region of essentially 100\% BEC, it corresponds to
  an equivalent surface density of about $10^{14}$ atoms/cm$^2$.

   Whatever the density is, the kinetic energy functional of a
  Bose-condensed system always has a contribution directly related
to
  $\Phi({\bf r})$ which is given by
\protect\cite{alf-jdw71,pn-dp90}
  \begin{eqnarray}
  \int d{\bf r}\Phi^\ast({\bf r})[-
  \frac{\hbar^2\nabla^2}{2m}]\Phi({\bf r})
  &=&\int d{\bf r} \frac{\hbar^2}{2m}\vert\bbox{\nabla}
  \sqrt{\rho_c({\bf r})}\vert^2\nonumber\\
  & +&\frac{1}{2}m\int d{\bf r}
  \rho_c({\bf r})
  {\bf v}^2_s({\bf r})\ ,
  \label{eq15}
  \end{eqnarray}
  where $\Phi({\bf r})\equiv\sqrt{\rho_{c}({\bf r})} e^{iS({\bf
r})}$
  and $m{\bf v}_s({\bf r})\equiv\hbar \nabla S({\bf r})$ is the
local
superfluid
  velocity. The expression in (\protect\ref{eq15}) is the origin of
  the kinetic energy term in (\protect\ref{eq8}) and
  (\protect\ref{eq11}).

  The long-range coherence effects associated with
  superfluidity \protect\cite{pn-dp90} are related to the local
phase
  $S({\bf r})$ of the complex (two-component) Bose order parameter
  $\Phi({\bf r})$ defined
  above. In a situation like the one considered here, where the
local condensate density $\rho_c({\bf r})$ is rapidly decreasing
normal to the surface, the superfluid flow properties may be quite
different parallel and perpendicular to the surface. This clearly
has experimental implications.

  Phenomenological density functional theories used for
inhomogeneous
  superfluid $^4$He \protect\cite{ss-jt87,jdr-mh-etc,fd-al-etc}
  always start from an energy functional of $\rho({\bf r})$ alone,
usually
  of the form
  \begin{equation}
  H[\rho({\bf r})]=\int d{\bf r}
  \frac{\hbar^2}{2m}\vert\bbox{\nabla}\sqrt{\rho({\bf
  r})}\vert^2+V_{{\rm
  corr}}[\rho({\bf r})]\ .
  \label{eq16}
  \end{equation}
  The first term is the kinetic energy of an
  inhomogeneous ideal Bose gas with a given local density
  $\rho({\bf r})$, while $V_{{\rm corr}}[\rho({\bf r})]$ is a
  phenomenological expression used to describe the effects of
  interactions. More precisely, the first term in
  (\protect\ref{eq16}) describes the energy of a fictitious liquid
of
  $N$ atoms
  as if they all occupied the same single-particle state
$\phi_0({\bf
  r})=\sqrt{\rho({\bf r})/N}$, corresponding to $\Phi({\bf
  r})=\sqrt{\rho({\bf r})}$.
  A complete density functional theory of Bose-condensed liquids
would involve
  functionals of both the local density $\rho({\bf r})$ and the
local order
  parameter $\Phi({\bf r})$ \protect\cite{noteag}. In such a
theory, the
  first term of (\protect\ref{eq15}) would naturally arise in place
of
  the first term in  (\protect\ref{eq16}), although we note that
the two
  terms coincide in the important low density surface region since
$\rho_c({\bf r}) \simeq \rho({\bf r})$.

  It would be useful to have finite temperature path
  integral
  Monte Carlo calculations of $\rho_{c}({\bf r})$
  in the surface region of superfluid $^4$He \protect\cite{dmc}.
  Physically, it seems
  clear that the surface region will Bose condense at the bulk
  superfluid transition temperature $T_\lambda\simeq 2.17$K.  The
  chemical potential is
  the same for the liquid and vapor in equilibrium and since it is
  large and negative $(\mu\simeq-7$K from 0 to 2K), the $^4$He
vapor
  can always be described as an ideal classical gas.
  Fig.~\protect\ref{fig1} is only valid at low
  temperatures ($\lesssim 1$K). As the
  temperature increases, the condensate density $\rho_c({\bf r})$
  will be increasingly depleted \protect\cite{ag93} and thus it
will
  be a
  smaller component of
  the total local density $\rho({\bf r})$, in {\it both} the bulk
and
  surface regions. However we expect that $\rho_c({\bf r})$ will
  still decrease exponentially in the surface region,
  while $\rho({\bf r})$ will go over smoothly
  to the small but finite density of the $^4$He vapor.

  In conclusion, in this letter we have argued that the surface
  region of superfluid $^4$He in self-bound phases can be described
  as an inhomogeneous dilute Bose gas which is completely
  Bose-condensed at $T=0$.  Dramatic evidence for this ``self-bound
  Bose
  gas'' is given by the numerical calculations in
  Ref.~\protect\cite
  {dsl-vkp-scp88}. We
  have noted that the asymptotic low
  density surface region can be described {\it exactly} by the
order
  parameter $\Phi({\bf r})$ given by the solution of a generalized
  Gross-Pitaevskii
  equation (\protect\ref{eq11}) taking full account of the long
  range van der Waals tail of the He--He potential. Optical
excitation of the surface atoms might be a way of detecting the
completely Bose-condensed nature of the surface region of
superfluid $^4$He.  We
  have used our new physical ideas to suggest a more microscopic
  basis for the kinetic energy term used in density
  functional theories.  This inhomogeneous
  Bose-condensed gas is readily available, and should
  complement the more exotic Bose gas systems on which much current
  research has
  concentrated \protect\cite{newref29}.

We have had useful discussions concerning these results with
F.~Lalo\"e and P.~Nozi\`eres.   A.G. would like to thank the
Universit\`{a} di Trento for
  financial support and stimulating hospitality during a sabbatical
  leave.  His
  research was also supported by NSERC of Canada.

 \begin{figure}
  \caption{The total density $\rho (r)$ and condensate
 density $\rho_c(r)$ as a function of the distance from the center
  of a $^4$He droplet of 70 atoms, based on Fig.~5 of
  Ref.~\protect\cite{dsl-vkp-scp88}.  For simplicity, we
  have smoothed out the small oscillations in $\rho(r)$.}
  \label{fig1}
  \end{figure}




\begin{references}

  \bibitem[*]{ag}Permanent address : Department
  of Physics, University of Toronto, Toronto, Ontario, Canada, M5S
1A7.

  \bibitem{ag-ds-ss95}{\bf Bose-Einstein Condensation}, ed.\ by
  A.~Griffin, D.~Snoke and S.~Stringari (Cambridge, N.Y., 1995).

  \bibitem{gt94}For a review, see G.~Taubes, Science {\bf 265}, 184
  (1994).

  \bibitem{ag93}A.~Griffin, {\bf Excitations in a Bose-Condensed
  Liquid} (Cambridge, N.Y., 1993).

  \bibitem{fd-al-etc}F.~Dalfovo, A.~Lastri, L.~Pricaupenko,
  S.~Stringari and J.~Treiner,  Phys.\ Rev.\ {\bf B52}, 1193
(1995).

  \bibitem{alf-jdw71}A.L.~Fetter and J.D. Walecka, {\bf Quantum
  Theory of Many-Particle Systems} (McGraw-Hill, N.Y., 1971),
  p.~488ff.

   \bibitem{pch-pcm65} P.C.~Hohenberg and P.C.~Martin, Annals of
Physics
  {\bf 34}, 291 (1965).


  \bibitem{dsl-vkp-scp88}D.S.~Lewart, V.R.~Pandharipande and
  S.C.~Pieper, Phys.~Rev.~{\bf B37}, 4950 (1988).

  \bibitem{vrp-scp-rb86}The variational wavefunction $\Psi_v$ used
in
  Ref.\protect\cite{dsl-vkp-scp88} is
  discussed in more detail in V.R.~Pandharipande, S.C.~Pieper and
  R.B.~Wiringa, Phys.~Rev.~{\bf B34}, 4571 (1986).

  \bibitem{sac93}S.A.~Chin, J.~Low Temp.~Phys.~{\bf 93}, 921
  (1993).

  \bibitem{ek85}E.~Krotscheck, Phys.~Rev.~{\bf B32}, 5713 (1985).

  \bibitem{cec93}C.E.~Campbell, J.~Low Temp.\ Phys.~{\bf 93}, 907
  (1993). The ``surface condensate'' discussed here
  concerns a new kind of surface ODLRO which  might arise in
  finite-sized droplets (an eigenvalue of O($N^{ 2 /  3 }$)
  instead
  of O($N$)). It is not related to the phenomenon discussed in the
  present letter.

  \bibitem{ek-gxq-wk85}See, for example, E.~Krotscheck, G.-X.~Qian
  and W.~Kohn,
  Phys.\ Rev.~{\bf B31}, 4245 (1985); S.A.~Chin and E.~Krotscheck,
  Phys.\ Rev.\ {\bf B45}, 852 (1992); K.A.~Gernoth, J.W.~Clark,
  G.~Senger and M.L.~Ristig, Phys.\ Rev.\ {\bf B49}, 15836 (1994).


  \bibitem{ss91}S.~Stringari, J.~Low Temp.~Phys.~{\bf 84}, 279
  (1991); Phys.~Rev.~{\bf B46}, 2974 (1992).

  \bibitem{rmb70}R.M.~Bowley, J.~Phys.~{\bf C3}, 2012 (1970).


  \bibitem{ccc-mc73}C.C.~Chang and M.~Cohen, Phys.~Rev.~{\bf A8},
  3131 (1973).

  \bibitem{ksl-mhk-gvc75}K.S.~Liu, M.H.~Kalos and G.V.~Chester,
  Phys.~Rev.~{\bf B12}, 1715 (1975).

  \bibitem{ss-jt87}S.~Stringari and J.~Treiner, Phys.~Rev.~{\bf
B36},
  8369 (1987); J.~Chem.~Phys.\ {\bf 87}, 5021 (1987).

  \bibitem{jdr-mh-etc}J.~Dupont-Roc, M.~Himbert, N.~Pavloff and
  J.~Treiner, J.~Low Temp.~Phys.~{\bf 81}, 31 (1990).


  \bibitem{pn-dp90}P.~Nozi\`{e}res and D. Pines, {\bf Theory of
  Quantum Liquids}, Vol.~II (Addison-Wesley, Redwood City,
  1990), Ch.~10.

  \bibitem{notemu} See Eqs.(4) and (8) of first paper in
  Ref.\protect\cite{ss91}.

  \bibitem{op-lo56} O.~Penrose and L.~Onsager, Phys.~Rev.~{\bf
104},
  576 (1956).

  \bibitem{epg61}L.P.~Pitaevskii, Sov.~Phys.- JETP {\bf 13}, 451
  (1961); E.P.~Gross, Nuovo Cimento {\bf 20}, 454 (1961).

\bibitem{newref23}One can prove that the correlation function in
(\protect\ref{eq8}) is equal to $\rho({\bf r}^\prime)\Phi({\bf
r})[1+F_1({\bf r},{\bf r}^\prime)]$, where the function $F_1$ takes
into account all short-range correlations left out in
(\protect\ref{eq9}).   Using this result,
one can easily see that (\protect\ref{eq8}) can be written quite
generally in a form identical to (\protect\ref{eq11}) but with an
inhomogeneous ``screened'' potential $v_{eff}({\bf r},{\bf
r}^\prime)\equiv v({\bf r}-{\bf r}^\prime)[1+F_1({\bf r},{\bf
r}^\prime)]$. This formulation provides a rigorous basis for
determining $\Phi({\bf r})$ in regions where the density is
non-negligible (see also Ref.~\protect\cite{cec93}). In homogeneous
fluids, the function $F_1$ has been extensively
discussed by M.L.~Ristig and J.W.~Clark, Phys.\ Rev.\ {\bf B40},
4355 (1989) and Stringari \protect\cite{ss91}.

  \bibitem{ibm-doe79}See, for example, I.B.~Mantz and D.O.~Edwards,
  Phys.\
  Rev.\ {\bf B20}, 4518 (1979).

  \bibitem{notetr} The leading order result $\rho(z) \sim e^{-2Az}$
  obtained here was first suggested  by W.F.~Saam, Phys.\ Rev.\
{\bf A4}, 1278
  (1971) and T.~Regge, J. Low Temp. Phys. {\bf 9}, 123 (1972).

  \bibitem{notess} See Ref.~16 of E. Zaremba and W. Kohn,
Phys.~Rev.\ {\bf B15}, 1769 (1977).

  \bibitem{noteag}  A.~Griffin, Can.~Journ.~Phys., in press.

  \bibitem{dmc}P.~Sindzingre, M.L.~Klein and D.M.~Ceperley,
  Phys.~Rev.~Lett.\ {\bf 63}, 160 (1989).

\bibitem{newref29}Since this paper was submitted, essentially
complete BEC has been achieved in a dilute gas of Rb atoms trapped
in a harmonic potential well.  See M.H.~Anderson, J.R.~Ensher,
M.R.~Matthews, C.E.~Wieman and E.A.~Cornell, Science {\bf 269}, 198
(1995).
  \end{references}
\end{document}